\documentclass[a4paper,12pt]{elsarticle}
\usepackage{amssymb,amsmath,mathrsfs,bm}
\usepackage[breaklinks=true]{hyperref}
\usepackage{graphicx,fancyhdr,verbatim}
\usepackage{geometry,color,multirow,colortbl}
\usepackage[usenames, dvipsnames]{xcolor}
\usepackage{psfrag,overpic,tikz,breakcites}
\usepackage[export]{adjustbox}
\usepackage[mathlines,pagewise]{lineno}

\graphicspath{{./FIGURES/}}

\hypersetup{
    colorlinks = true,
    urlcolor   = black,
    citecolor  = black,
}
\newcommand\Rey{\mbox{\textit{Re}}}  
\newcommand\Pran{\mbox{\textit{Pr}}}  
\newcommand\Nu{\mbox{\textit{Nu}}}  
\newcommand\St{\mbox{\textit{St}}}  
\newcommand\Pe{\mbox{\textit{Pe}}}  
\newcommand\diff{\mbox{d}}  

\definecolor{darkviolet}{rgb}{0.58, 0.0, 0.83}
\definecolor{shamrockgreen}{rgb}{0.0, 0.62, 0.38}
\definecolor{deepskyblue}{rgb}{0.0, 0.75, 1.0}
\definecolor{amber}{rgb}{1.0, 0.49, 0.0}
\definecolor{red}{rgb}{1.0, 0.0, 0.0}
\definecolor{blue}{rgb}{0.0, 0.0, 1.0}
\definecolor{airforceblue}{rgb}{0.36, 0.54, 0.66}
\definecolor{amethyst}{rgb}{0.6, 0.4, 0.8}

\newcommand\solid[1][1.0cm]{\rule[0.4ex]{#1}{2.8pt}}

\modulolinenumbers[5]
\begin{document}

\begin{frontmatter}

	\title{Mean temperature profiles in \\ turbulent internal flows}
\date{\today}

\author[add1]{Sergio Pirozzoli}
\author[add2]{Davide Modesti}

\address[add1]{Dipartimento di Ingegneria Meccanica e Aerospaziale, 
Sapienza Universit\'a di Roma, via Eudossiana 18, 00184 Roma, Italia}
	\address[add2]{Faculty of Aerospace Engineering, Delft University of Technology, Kluyverweg 2, 2629 HS Delft, The Netherlands}

\begin{abstract}
We derive explicit formulas for the mean profiles 
of temperature (modeled as a passive scalar) in forced turbulent convection, as a function of the Reynolds and Prandtl numbers.
The derivation leverages on the observed universality of the inner-layer 
thermal eddy diffusivity with respect to Reynolds and Prandtl number variations
and across different flows, and on universality of the passive scalar defect in the core flow.
Matching of the inner- and outer-layer expression yields a smooth compound mean temperature profile.
We find excellent agreement of the analytical profile with data from direct numerical simulations of pipe and channel flows under
various thermal forcing conditions, and over a wide range of Reynolds and Prandtl numbers.
\end{abstract}


\end{frontmatter}


\section{Introduction}

The exploration of passive scalars within turbulent flows bounded by walls holds significant practical implications. It plays a crucial role in comprehending the behavior of diluted contaminants and serves as a model for temperature distribution assuming low Mach numbers and minimal temperature disparities \citep{monin_71, cebeci_84}. Nonetheless, quantifying 
minute temperature differences and the concentration of passive tracers
poses formidable challenges, leading to restricted insights into fundamental passive scalar statistics~\citep{gowen_67, kader_81, subramanian_81, nagano_88}.

The investigation of passive scalars in turbulent flows predominantly centers on scenarios where the Prandtl number ($\mathrm{Pr}$) approaches unity, representing the ratio of kinematic viscosity to thermal diffusivity ($\mathrm{Pr}=\nu/\alpha$). Numerous studies have discussed the close similarities between the passive scalar field and the streamwise velocity field under these conditions \citep{kim_87, abe_09, antonia_09}. However, various fluids, including water, engine oils, glycerol, and polymer melts, exhibit Prandtl numbers significantly exceeding unity, while liquid metals and molten salts have markedly lower Prandtl numbers.

For the diffusion of contaminants, the role of the Prandtl number is replaced by the Schmidt number, which denotes the ratio of kinematic viscosity to mass diffusivity. In practical applications, the Schmidt number typically far exceeds unity \citep{levich_62}. In such instances, the resemblance between velocity and passive scalar fluctuations is significantly compromised, rendering predictions of even fundamental flow properties notably challenging.

Regarding wall fluxes, the most robust framework established to date is attributed to the work by \citet{kader_72}. Drawing upon universality arguments, these authors derived a predictive law for the nondimensional flux (Nusselt number) as a function of the Prandtl number. This framework primarily involves modeling the logarithmic offset function, which represents the Prandtl-dependent additive constant in the overlap-layer mean passive scalar profiles. Despite the solidity of this framework, semi-empirical power-law correlations \citep{dittus_33, kays_80} continue to find widespread usage in engineering design.
As for the mean profiles of passive scalars, the most comprehensive study available can be traced back to the work of \citet{kader_81}. In this study, an empirical interpolation formula was derived, connecting the universal near-wall conductive layer with the outer logarithmic layer. This interpolation formula was observed to reasonably align with the temperature profile behavior observed in experiments available at the time.

\citet{pirozzoli_23c} studied the statistics of passive scalars in pipe flow
in the range of Prandtl numbers from 
$\Pran=0.00625$ to $\Pran=16$, using direct numerical simulation (DNS) of the Navier--Stokes equations, 
and introduced an eddy-viscosity model to obtain fully explicit predictions of the mean passive scalar
profiles in the inner layer, and for the corresponding logarithmic offset function. 
Asymptotic scaling formulas
were also derived for the thickness of the diffusive sub-layer, and for the heat transfer coefficient,
which were found to accurately represent variations of both the Reynolds and the Prandtl number, 
for $\Pran \gtrsim 0.00625$. A similar study of turbulent heat transfer in plane channel flows 
based on DNS data was carried out by \citet{pirozzoli_23a}, who also reported predictive formulas 
for the heat transfer coefficients under various thermal forcing conditions. 
In this paper, we use the pipe and channel flow DNS database to
derive fully explicit analytical representations of the mean passive scalar profiles 
throughout the wall layer including the wake region, as a function of the Reynolds and Prandtl numbers,
thus expanding on the work of \citet{pirozzoli_23c}.
This information is of outstanding practical importance as it can be used 
as a benchmark for low-order fidelity models like RANS, and it allows
implementation of techniques of modal analysis~\citep{taira_17}.
Although, as previously pointed out, the study of passive scalars is relevant 
in several contexts, one of the primary fields of application is heat transfer, and
therefore from now on we will refer to the passive scalar field as 
the temperature field (denoted as $T$), and passive scalar fluxes will be interpreted as heat fluxes.

\section{Predictive formulas}

\subsection{The inner layer}

The starting point for the analysis of the mean temperature profile in the inner layer
is the mean thermal balance equation, which in internal flows reads 
\begin{equation}
\frac 1{\Pran} \frac{\diff \Theta^+}{\diff y^+} 
-\left< v \theta \right>^+   
= 1 - \eta,  \label{eq:mtb}
\end{equation}
where $\theta = T-T_w$ expresses the temperature difference with respect to the wall, $v$ is the wall-normal velocity,
$\eta = y/\delta_t$, with $y$ the wall distance, and $\delta_t$ a suitable measure of the thickness
of the thermal layer, to be identified from case to case as later explained.
Here and in the following, capital symbols are used to denote Reynolds averaged quantities, 
whereas lowercase symbols are used to denote fluctuations thereof, and angle brackets 
denote the averaging operator.
In equation~\eqref{eq:mtb} the $+$ superscript denotes normalization in wall units,
whereby the friction velocity ($u_{\tau} = (\tau_w/\rho)^{1/2}$) is used for velocities, 
the viscous length scale ($\delta_v = \nu / u_{\tau}$) is used for lengths, and 
the friction temperature ($T_{\tau}=\alpha/u_{\tau} \left(\diff \langle T \rangle/\diff y\right)_w$) is used for temperature.
The mean wall shear stress is $\tau_w$, and $\rho$ and $\nu$ and the fluid density and viscosity, respectively. 

Modeling the turbulent heat flux $\left< v \theta \right>$ in \eqref{eq:mtb} 
requires closure with respect to the mean temperature gradient~\citep[see, e.g.][]{cebeci_84},
through the introduction of a thermal eddy diffusivity, defined as
\begin{equation}
\alpha_t = \frac{-\left< v \theta \right>}{\diff \Theta / \diff y}. \label{eq:alphat}
\end{equation}
	Asymptotic consistency at the wall~\citep{kader_72} requires that the turbulent flux scales as 
$-\left< v \theta \right>^+ \sim y^3$, and equation~\eqref{eq:mtb} implies 
$\frac{\diff \Theta^+}{\diff y^+} \approx \Pran + O({y^+}^2/\Rey_{\tau})$.
Hence, the leading-order behavior of the thermal eddy diffusivity at the wall is
\begin{equation}
\alpha_t^+ = \frac{-\left< v \theta \right>^+}{\diff \Theta^+ / \diff y^+} \sim {y^+}^3. \label{eq:alphat_wall}
\end{equation}

As noted by \citet{pirozzoli_23c}, the thermal eddy diffusivity in the inner layer of 
wall-bounded flows has a relatively simple behavior, and it is very nearly universal 
with respect to variations of the Reynolds number.
The thermal eddy diffusivity is also very much insensitive to variations of the Prandtl number, 
with exception of vanishingly small Prandtl numbers, in which limit $\alpha_t$
must vanish as conduction takes over.
This is well portrayed in figures~\ref{fig:alphat_pipe} and \ref{fig:alphat_chan},
which we will comment later on.
The occurrence of a logarithmic layer in the mean temperature profile implies
$\alpha_t \sim y$ away from the wall, hence \citet{pirozzoli_23c}
suggested the following functional expression to model the 
thermal eddy viscosity throughout the inner layer
\begin{equation}
\alpha_t^+ = \frac{(k_{\theta} {y^+})^3}{(k_{\theta} {y^+})^2+C_{\theta}^2}. \label{eq:musker}
\end{equation}
Equation~\eqref{eq:musker} is inspired by the work of \citet{musker_79}, who used a similar function to model
the eddy viscosity in turbulent boundary layers.
In equation~\eqref{eq:musker} the thermal K\'arm\'an constant  was determined to be 
$k_{\theta} \approx 0.459$~\citep{pirozzoli_22}, 
and $C_{\theta}=10.0$ was found based on scrutiny of DNS data for pipe flow.
Whereas alternative functional expressions are possible,
equation~\eqref{eq:musker} bears the substantial advantage of being amenable to further analytical developments.

Starting from equation~\eqref{eq:mtb}
and under the near-wall approximation ($\eta << 1$), 
one can infer the distribution of the mean temperature in the inner layer 
from knowledge of the eddy thermal diffusivity, by integrating
\begin{equation}
\frac{\diff \Theta^+}{\diff y^+} = \frac {\Pran}{1 + \Pran \, \alpha_t^+} , \label{eq:dTdy}
\end{equation}
with $\alpha_t$ given in equation~\eqref{eq:musker}.
The result of the integration yields the mean temperature profile in the inner layer
\begin{equation}
 \begin{split}
\Theta_i^+ (y^+, \Pran) = \frac{1}{2 k_{\theta} \zeta_0 (2 + 3 \Pran \zeta_0)}
\left\{
\frac {2\left( 2 \zeta_0 + 3 \Pran^2 C_{\theta}^2 \zeta_0 + \Pran (C_{\theta}^2+2 \zeta_0^2) \right)}{\Delta} 
\right. \\
\times \left[ 
\arctan{\left( \frac{1+\Pran \zeta_0}{\Delta} \right)} \right. 
-
\left. \arctan{\left( \frac{1+\Pran (2 \zeta + \zeta_0)}{\Delta} \right)}
\right] \\
+ 2 \Pran \left( C^2 + \zeta_0^2 \right) \log (1-\frac{\zeta}{\zeta_0}) \\
+ 
\left. 
\vphantom{\frac {2\left( 2 \zeta_0 + 3 \Pran^2 C_{\theta}^2 \zeta_0 + \Pran (C_{\theta}^2+2 \zeta_0^2) \right)}{\Delta}}
\left( \Pran (2 \zeta_0^2 - C_{\theta}^2 ) + 2 \zeta_0 \right) \log{\frac{\Pran \zeta^2 + (1 + \Pran \zeta_0)(\zeta+\zeta_0)}{\zeta_0 (1 + \Pran \zeta_0)}} 
\right\} ,
 \end{split}
 \label{eq:Tprof}
\end{equation}
where $\zeta=k_{\theta} y^+$, $\Delta = (3 \Pran^2 \zeta_0^2 + 2 \Pran \zeta_0 -1)^{1/2}$, and $\zeta_0$ is the single (negative) real root of the cubic equation
\begin{equation}
\Pran \zeta^3 + \zeta^2 + C_{\theta}^2 = 0, 
\label{eq:cubic}
\end{equation}
whose exact solution is
\begin{equation}
\zeta_0 = \frac 1{3 \Pran} \left( - 1 + \frac 1z + z \right), z= \left[ \frac 12 \left( -2 - 27 \Pran^2 C_{\theta}^2 + \sqrt{-4+ \left( 2 + 27 \Pran^2 C_{\theta}^2 \right)^2} \right) \right]^{1/3} . \label{eq:cardano}
\end{equation}

The temperature profiles given in equation~\eqref{eq:Tprof} exhibit
logarithmic behaviour at $y^+ >> 1$, namely
\begin{equation}
\Theta^+ = \frac 1{k_{\theta}} \log y^+ + \beta (\Pran), \label{eq:loglaw}
\end{equation}
where the log-law offset function is given by the following asymptotic expression
\begin{equation}
\beta(\Pran) 
= \frac 1{k_{\theta}} \left[ \frac{2 \pi C_{\theta}^{2/3}}{3 \sqrt{3}} \Pran^{2/3} 
+ \frac 13 \log \Pran - \left( \frac 16 + \frac 1{2 \sqrt{3}} + \frac 23 \log C_{\theta} - \log k_{\theta} \right) \right] + O(\Pran^{-2/3}). \label{eq:beta_highPr}
\end{equation}

Other synthetic profiles for the mean temperature in the inner layer are available in the literature. A notable example is the
approach described by~\citet{cebeci_84}, where the authors used the eddy-diffusivity and eddy-viscosity formulas
from classical turbulence models to derive temperature profiles at arbitrary Prandtl numbers.
Another example is the empirical correlations based on experimental-data fitting proposed by \citet{kader_81}, which we
use as reference in study.
The main advantage of the present framework compared to existing formulas is that equation~\eqref{eq:Tprof} provides an explicit analytical expression
for the mean temperature profiles based on a simple expression for the eddy diffusivity, which embeds the correct asymptotic 
behavior close to the wall and in the logarithmic layer.

\subsection{The core layer}

The behavior of the mean temperature in the core (wake) region of wall-bounded flows was
studied in terms of the temperature defect function by \citet{pirozzoli_16} and \citet{pirozzoli_21}.
The key finding was that the temperature defect profile (with respect to the peak value)
is very nearly universal with respect to both Reynolds and Prandtl number variations, 
and the universal region encompasses a wide part of the flow thickness.
Departures from outer-layer universality were only observed at 
$\Pran \lesssim 0.025$, below which 
the similarity region becomes progressively confined to the outermost part of the thermal wall layer.
As suggested by \citet{pirozzoli_16,pirozzoli_22}, the core temperature 
profiles in internal flows can be closely approximated with simple universal quadratic distributions, 
which one can derive from \eqref{eq:mtb} under the assumption 
of constant eddy thermal diffusivity. This assumption
is analogous to the hypothesis of constant eddy viscosity proposed by~\citet{clauser_56},
also leading to a parabolic outer layer velocity distribution.
A convenient analytical representation for the core mean temperature profile $\Theta_c$ is then
\begin{equation}
\Theta_{_{e}}^+ - \Theta_c^+ = C_w \left( 1 - \eta \right)^2, \label{eq:parabola}
\end{equation}
where $\Theta_{_{e}}$ is the value of the mean temperature at the edge of the thermal wall layer.

\subsection{Patching} \label{sec:matching}

An explicit approximation for the mean temperature profile throughout the 
thermal wall layer is then obtained by patching the inner-layer profile \eqref{eq:Tprof} 
with the core profile ~\eqref{eq:parabola}. Smooth patching is obtained by
setting the transition point at the wall distance where the 
asymptotic logarithmic profile \eqref{eq:loglaw} and the core temperature profile~\eqref{eq:parabola}
match and have equal first derivative. It is easy to show that continuity of the 
first derivatives is achieved provided the patching point is placed at the coordinate
\begin{equation}
\eta^* = \left. \left( 1-\sqrt{ 1 -\frac 2{C_w k_{\theta}}} \right) \right/ 2, 
\label{eq:etas}
\end{equation}
whereas matching the values of the inner-layer and core temperature profiles implies that
\begin{equation}
\Theta_e^+ = \Theta_i^+ (\eta^* \delta_t^+, \Pran) + C_w \left( 1 - \eta^* \right)^2 , \label{eq:ThetaE}
\end{equation}
to be used in equation~\eqref{eq:parabola}.



\begin{table}
 \centering
\begin{tabular*}{1.\textwidth}{@{\extracolsep{\fill}}lccccc}
	Flow & Heating & $\delta_t$ & $C_{\theta}$ & $C_w$ & $\eta^*$ \\
 \hline
	Pipe    & UIH      & $R$   & 10.0 & 6.00 & 0.238  \\ 
	Pipe    & CHF      & $R$   & 10.0 & 7.00 & 0.193  \\
	Channel & UIH-sym  & $h$   & 10.0 & 5.48 & 0.274  \\
	Channel & UIH-asym & $2 h$ & 10.0 & 12.3 & 0.0982  \\
 \hline
\end{tabular*}
\caption{Parameters for compound mean temperature profiles obtained by patching equation~\eqref{eq:Tprof} 
and \eqref{eq:parabola}. $\delta_t$ is the assumed thickness of the thermal layer, with $R$ the pipe radius and
$h$ the channel half-thickness. $C_{\theta}$ is the inner-layer constant to be used in \eqref{eq:musker}, $C_w$ is the core profile constant to use in \eqref{eq:parabola}, and $\eta^*$ is the outer-scaled coordinate of the patching point
between the inner and the core temperature profiles.}
\label{tab:params}
\end{table}

\section{Results}

\subsection{Pipe flow} \label{sec:pipe}

\begin{table}
 \centering
\begin{tabular*}{1.\textwidth}{@{\extracolsep{\fill}}lccccc}
	$\Pran$ & Mesh ($N_z \times N_r \times N_{\phi}$) & $\Pe_{\tau}$ & $\Nu_{UIH}$ & $\Nu_{CHF}$ & Line style \\
 \hline
 0.00625   & $1792 \times 164 \times 1793$ & 7.11    & $ 8.02  $ & 7.35 & \color{darkviolet}\solid \\
 0.0125    & $1792 \times 164 \times 1793$ & 14.2    & $ 9.41  $ & 8.68 & \color{amethyst}\solid \\
 0.025     & $1792 \times 164 \times 1793$ & 28.5    & $ 12.6  $ & 11.6 & \color{magenta}\solid \\
 0.0625    & $1792 \times 164 \times 1793$ & 71.1    & $ 21.5  $ & 20.2 & \color{lime}\solid \\
 0.125     & $1792 \times 164 \times 1793$ & 142.2   & $ 34.2  $ & 32.5 & \color{shamrockgreen}\solid \\
 0.25      & $1792 \times 164 \times 1793$ & 284.4   & $ 53.8  $ & 51.4 & \color{deepskyblue}\solid \\
 0.5       & $1792 \times 164 \times 1793$ & 568.8   & $ 81.7  $ & 79.0 & \color{airforceblue}\solid \\
 1         & $1792 \times 164 \times 1793$ & 1137.6  & $119.9  $ & 116.6 & \color{black}\solid \\
 2         & $3584 \times 269 \times 3584$ & 2275.2  & $168.0  $ & 165.0 & \color{amber}\solid \\
 4         & $3584 \times 269 \times 3584$ & 4550.4  & $233.3  $ & 229.7 & \color{blue}\solid \\
 16        & $7168 \times 441 \times 7168$ & 18201.6 & $421.2  $ & 419.4 & \color{red}\solid \\
 \hline
\end{tabular*}
\caption{Flow parameters for DNS of pipe flow at various Prandtl number. 
$N_z$, $N_r$, $N_{\phi}$ denote the number of grid points in the axial, radial, and
azimuthal directions, respectively; $\Pe_{\tau} = \Pran \, \Rey_{\tau}$ is the friction P\'eclet number;
$\Nu$ is the Nusselt number (as defined in equation~\eqref{eq:nusselt}), for cases with 
uniform internal heating (UIH), and constant heat flux (CHF).
All DNS are carried out in a computational domain with length $L_z = 15 R$, at
bulk Reynolds number $\Rey_b=44000$, corresponding to friction Reynolds number $\Rey_{\tau} \approx 1138$.}
\label{tab:runs_pipe}
\end{table}

Results of DNS of thermal pipe flow are reported for the two canonical cases
of uniform internal heating (UIH), and constant heat flux (CHF).
In the former case the energy equation is forced with a spatially uniform internal heating term,
in such a way that the bulk temperature is kept constant in time~\citep{kim_89,pirozzoli_16}.
In the latter case, the forcing term varies from point to point proportionally to
$u/u_b$, such that the bulk temperature is also constant~\citep{kawamura_99,abe_04,alcantara_21}.
This second approach more precisely mimics the physical case of thermally developed flow 
in a duct with spatially and temporally uniform wall heating~\citep{cebeci_84}.
As discussed by \citet{abe_04,alcantara_21}, the forcing strategy has small, but non-negligible
effect on the temperature statistics.
In both cases, the maximum temperature is attained at the pipe centreline, hence 
we assume the thermal wall layer thickness to be the pipe radius, and accordingly set $\delta_t = R$.

A wide range of Reynolds and Prandtl numbers has been explored, with 
spanning Prandtl numbers from 0.0025 to 16 and friction Reynolds numbers 
$\Rey_{\tau} (= R u_{\tau} / \nu)$ ranging
from $180$ to $6000$~\citep{pirozzoli_22, pirozzoli_23b}.
A full scan of Prandtl numbers is here reported, for fixed bulk Reynolds number $\Rey_b = 44000$,
corresponding to friction Reynolds number $\Rey_{\tau} \approx 1138$.
The corresponding computational parameters are reported for reference in table~\ref{tab:runs_pipe}.
In the table we also report some key global parameters as the friction P\'eclet number,
$\Pe_{\tau} = \Pran \Rey_{\tau}$, and the Nusselt number,
\begin{equation}
\Nu = \Rey_b \, \Pran \, \St , \label{eq:nusselt}
\end{equation}
with $\St$ the Stanton number, defined as
\begin{equation}
\St
= \frac{\alpha \left< \frac {\diff {T}}{\diff y} \right>_w}{u_b \left( T_m - T_w \right)} ,
\label{eq:stanton}
\end{equation}
where $u_b$ is the bulk velocity and $T_m$ is the mixed mean temperature~\citep{kays_80}.
The table supports the generally accepted notion that the specific forcing has little 
effect of the global thermal performance at Prandtl number of order unity or higher, 
however percent differences in the Nusselt number become significant at low Prandtl numbers.

\begin{figure}
 \centerline{
(a)~\includegraphics[width=7.0cm]{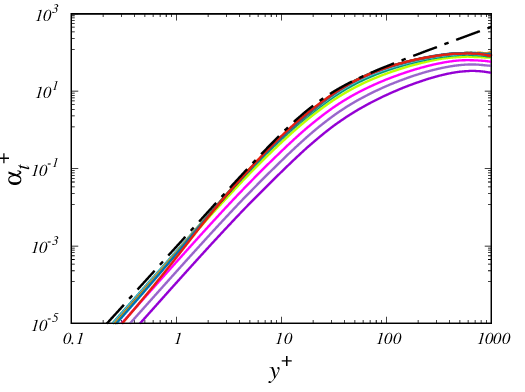}~(b)~\includegraphics[width=7.0cm]{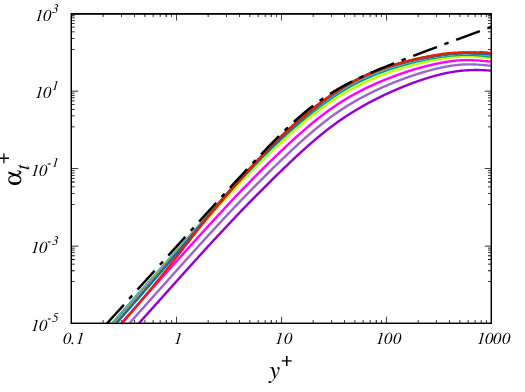}}
\caption{Distributions of inferred eddy thermal diffusivity ($\alpha_t$) as a function of wall distance 
for pipe flow with uniform internal heating (a) and constant heat flux (b), at various $\Pran$, for $\Rey_{\tau}=1138$.
The dash-dotted line denotes the fit given in equation~\eqref{eq:musker}.
Colour codes are as in table~\ref{tab:runs_pipe}.}
\label{fig:alphat_pipe}
\end{figure}

The inferred eddy thermal diffusivities for all cases at $\Rey_{\tau} = 1138$ are reported
in figure~\ref{fig:alphat_pipe}. The agreement with the fit given in equation~\eqref{eq:musker} is 
quite good in the inner layer, with deviations occurring only at $\Pran \lesssim 0.025$.
Slight deviations from the DNS are found at $y^+ \lesssim 10$, where in any case the eddy diffusivity 
is much less than the molecular one.
Notably, the influence of the thermal forcing (compare the two panels) seems to be entirely negligible.
Universality with respect to Reynolds number variations has also been verified, but figures are omitted.


\begin{figure}
 \centerline{
(a)~\includegraphics[width=7.0cm]{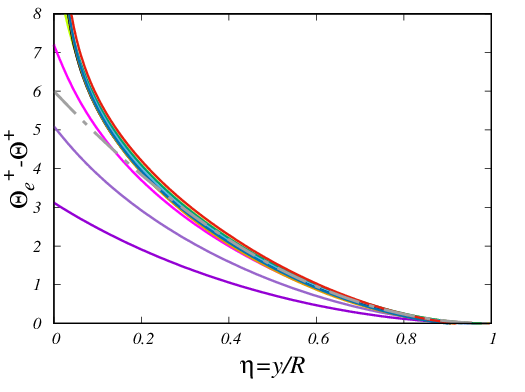}~(b)~\includegraphics[width=7.0cm]{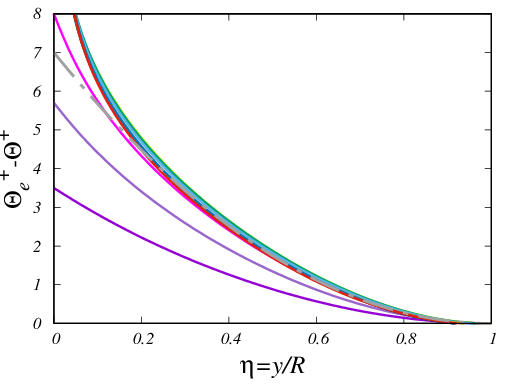}
 }
\caption{Mean temperature defect profiles 
for pipe flow with uniform internal heating (a) and constant heat flux (b), at various $\Pran$, for $\Rey_{\tau}=1138$.
The dot-dashed line marks a parabolic fit of the DNS data ($\Theta^+_{_{e}}-\Theta_c^+ = C_w (1-\eta)^2$), 
with values of $C_w$ given in table~\ref{tab:params}.
Colour codes are as in table~\ref{tab:runs_pipe}.}
\label{fig:defect_pipe}
\end{figure}

As for the temperature distribution in the core flow, 
in figure~\ref{fig:defect_pipe} we show the mean temperature profiles in
defect form, referred to the pipe centerline properties. Again, the temperature profiles at fixed $\Rey_{\tau}$ 
are shown at various Prandtl numbers. As claimed in previous studies~\citep{pirozzoli_23b}, 
the figure supports close universality of the defect temperature profiles, for $\eta = y/R \gtrsim 0.2$.
Furthermore, the figure confirms that the DNS data can be accurately fitted with the quadratic relationship~\eqref{eq:parabola}.
The fitting constant $C_w$, reported in table~\ref{tab:params} is a bit larger for the case of CHF forcing, on account
of a stronger temperature defect. 
The universality of defect temperature profiles concerning variations in Reynolds number has been corroborated in prior literature \citep{pirozzoli_22}, and is not reiterated here.

\begin{figure}
 \centerline{
 (a)~\includegraphics[width=7.0cm]{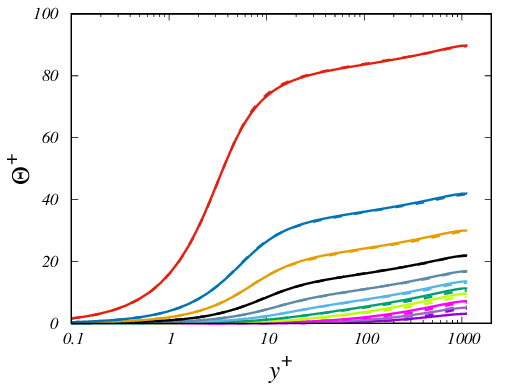}~(b)~\includegraphics[width=7.0cm]{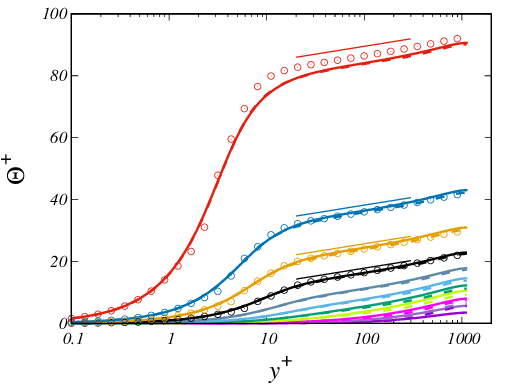}}
\caption{Comparison of mean temperature profiles obtained from DNS (solid lines) 
with the prediction of the compound fitting function given by \eqref{eq:Tprof} and \eqref{eq:parabola}, 
with fitting parameters listed in table~\ref{tab:params} (dashed lines),
for pipe flow with uniform internal heating (a) and with constant heat flux (b), at $\Rey_{\tau}=1138$.
In panel (b) the circles denote the empirical predictions put forth by \citet{kader_81},
and the thin lines the thermal law-of-the-wall of \citet{kays_80}.
Colour codes are as in table~\ref{tab:runs_pipe}.}
\label{fig:Tplus_pipe}
\end{figure}

Based on the observed universality of the inner and outer layers with respect to Reynolds and Prandtl 
number variations, we proceed to apply the patching procedure outlined in \S~\ref{sec:matching},
with fitting parameters listed in table~\ref{tab:params}.
The resulting temperature profiles are reported in figure~\ref{fig:Tplus_pipe}, which confirms that 
the quality of the resulting patched temperature 
profiles is generally very good. 
Visible deviations are only evident at extremely low Prandtl numbers, as anticipated due to the absence of a genuine overlap layer. Discrepancies from the predicted trends are observed specifically at the lowest Prandtl numbers, which, as previously noted, deviate from the universal trend of $\alpha_t$. 
In panel (b), we additionally present a comparison with the thermal law-of-the-wall proposed by \citet[][Eqn.(14-4)]{kays_80} and the empirical correlation proposed by \citet{kader_81}. To maintain clarity, only cases with $\Pran \geq 1$ are depicted. The figure purports substantial overprediction of $\Theta^+$ from the log-law formula of \citet{kays_80}. As for Kader's correlation, the general quality of the fit is more satisfactory, however an anomalous behavior is noticed in the buffer layer, and the shift in the log-law is somewhat overestimated at high $\Pran$.

\subsection{Channel flow}

\begin{table}
 \centering
\begin{tabular*}{1.\textwidth}{@{\extracolsep{\fill}}lccccccc}
	$\Pran$ & Mesh ($N_x \times N_y \times N_z$) & $\Pe_{\tau}$ & $\Nu_{SYM}$ & $\Nu_{ASYM}$  & Line style \\
 \hline
$0.025$     & $1536 \times 298 \times 2304$ & $24.7 $   & 10.0  & 5.88  & \color{magenta}\solid \\
$0.25$      & $1536 \times 298 \times 2304$ & $247.3$   & 44.6  & 32.7  & \color{deepskyblue}\solid \\
$0.5$       & $1536 \times 298 \times 2304$ & $494.4$   & 68.5  & 53.3  & \color{airforceblue}\solid \\
$1$         & $1536 \times 298 \times 2304$ & $1002.1$  & 101.7 & 86.3  & \color{black}\solid \\
$2$         & $3072 \times 485 \times 4608$ & $2010.4$  & 148.7 & 128.9 & \color{amber}\solid \\
$4$         & $3072 \times 485 \times 4608$ & $4019.6$  & 207.9 & 187.8 & \color{blue}\solid \\
\hline
\end{tabular*}
\caption{Flow parameters for DNS of channel flow at various Prandtl number. 
$N_x$, $N_y$, $N_z$ denote the number of grid points in the streamwise, wall-normal, and
spanwise directions, respectively; $\Pe_{\tau} = \Pran \, \Rey_{\tau}$ is the friction P\'eclet number;
$\Nu$ is the Nusselt number (as defined in equation~\eqref{eq:nusselt}), for cases with 
symmetric heating (SYM), and asymmetric heating with one adiabatic wall (ASYM).
All DNS are carried out in a computational domain with size $6 \pi h \times 2 h \times 2 \pi h$, at
bulk Reynolds number $\Rey_b=40000$, corresponding to friction Reynolds number $\Rey_{\tau} \approx 1002$.}
\label{tab:runs_chan}
\end{table}

A similar analysis is herein reported for thermal channel flow, based on the 
DNS data of \citet{pirozzoli_23a}.
Two cases of thermal forcing are considered with uniform internal 
heating, whereby the two walls of the channel are either kept at the same temperature,
hence heat is allowed to flow away through both walls, or
one of the two walls is kept adiabatic.
In the first case, referred to as symmetric heating (SYM),
the maximum temperature is attained at the channel centreline,
hence we assume the thermal wall layer thickness to be the channel half-thickness, $\delta_t=h$.
In the second case, referred to as asymmetric heating (ASYM), 
the maximum temperature is attained at the adiabatic wall,
hence we assume the thermal wall layer thickness to be the full channel thickness, $\delta_t=2 h$.
Just as for pipe flow, a relatively wide range of Reynolds 
and Prandtl numbers has been explored, with $\Rey_{\tau} (= h u_{\tau} / \nu)$ from 180 to 2000,
and $\Pran$ from 0.025 to 4.
A full scan of Prandtl numbers is here reported, for fixed bulk Reynolds number $\Rey_b (= 2 h u_b / \nu) = 40000$,
corresponding to friction Reynolds number $\Rey_{\tau} \approx 1002$.
The key parameters for the DNS of channel flow are listed in table~\ref{tab:runs_chan}.

\begin{figure}
 \centerline{
 (a)~\includegraphics[width=7.0cm]{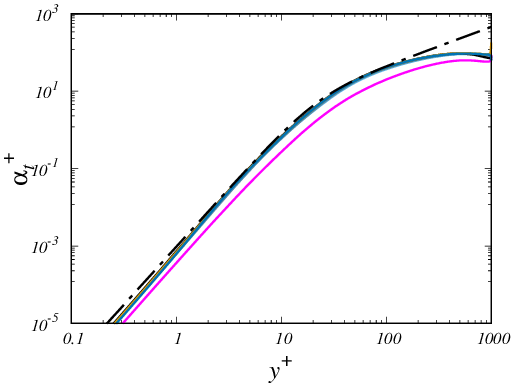}~(b)~\includegraphics[width=7.0cm]{fig4a}}
\caption{Distributions of inferred eddy thermal diffusivity ($\alpha_t$) as a function of wall distance,
for plane channel flow with symmetric heating (a), and with asymmetric heating (b), at various $\Pran$, for $\Rey_{\tau}=1002$.
The dash-dotted line denotes the fit given in equation~\eqref{eq:musker}.
Colour codes are as in table~\ref{tab:runs_chan}.}
\label{fig:alphat_chan}
\end{figure}

Figure~\ref{fig:alphat_chan} confirms that the inner-scaled thermal eddy diffusivity is universal
across the range of Prandtl numbers under scrutiny, the only outlier being the case
at the lowest $\Pran$. In agreement with pipe flow, we find that 
equation~\eqref{eq:musker} provides an excellent fit of the DNS data in the inner layer,
corroborating universality of the model for the temperature log-law shift function
introduced by \citet{pirozzoli_23c}, and whose validity was also proved for turbulent boundary layers~\citep{balasubramanian_23}.

\begin{figure}
 \centerline{
(a)~\includegraphics[width=7.0cm]{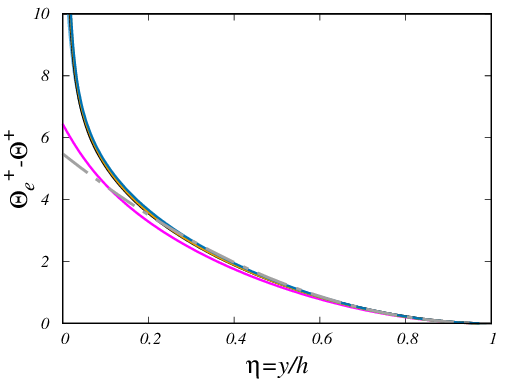}~(b)~\includegraphics[width=7.0cm]{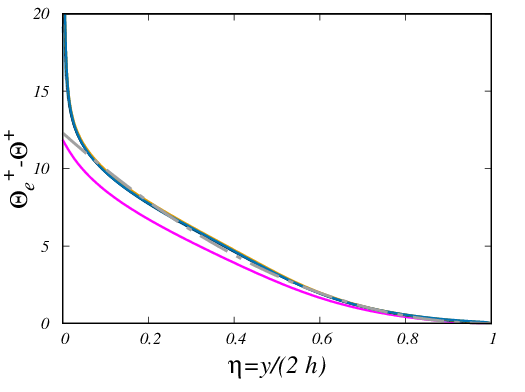}
 }
\caption{Mean temperature defect profiles 
for channel flow with symmetric heating (a), and with asymmetric heating (b), at various $\Pran$, for $\Rey_{\tau}=1002$.
The dot-dashed line marks a parabolic fit of the DNS data ($\Theta^+_{_{e}}-\Theta_c^+ = C_w (1-\eta)^2$), 
with values of $C_w$ given in table~\ref{tab:params}.
Colour codes are as in table~\ref{tab:runs_chan}.}
\label{fig:defect_chan}
\end{figure}

The mean temperature defect profiles are shown in figure~\ref{fig:defect_chan}.
In this respect, we note that the reference value is the mean temperature at the
channel centreline in the case with symmetric heating (left panel),
and the mean temperature at the adiabatic wall in the case of asymmetric heating (right panel).
Just like the case of pipe flow, the simple quadratic fit~\eqref{eq:parabola}, with suitably adjusted 
wake strength constants as given in table~\ref{tab:params} yields an
excellent approximation for the temperature defect, the only outlier being the case at the lowest $\Pran$.
It is especially noteworthy that in the case of asymmetric heating the quadratic approximation fits accurately the DNS
data in about $90\%$ of the channel thickness.

\begin{figure}
 \centerline{
 (a)~\includegraphics[width=7.0cm]{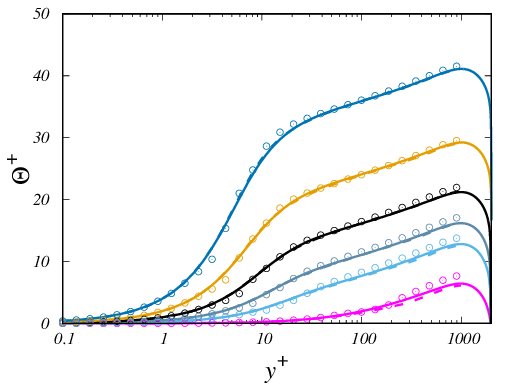}~(b)~\includegraphics[width=7.0cm]{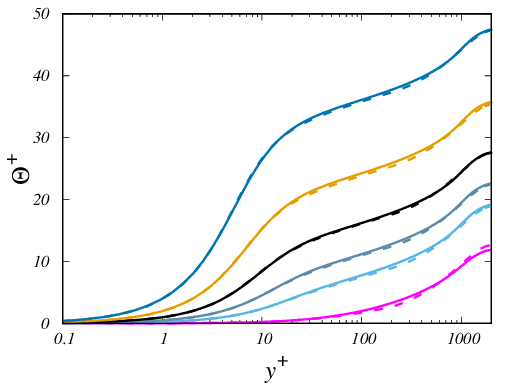}}
\caption{Comparison of mean temperature profiles obtained from DNS (solid lines) 
with the prediction of the compound fitting function given by \eqref{eq:Tprof} and \eqref{eq:parabola}, 
with fitting parameters listed in table~\ref{tab:params} (dashed lines),
for channel flow with symmetric heating (a) and with asymmetric heating (b), at $\Rey_{\tau}=1002$.
The circles depicted in panel (b) represent the empirical predictions put forth by \citet{kader_81}.
Colour codes are as in table~\ref{tab:runs_chan}.}
\label{fig:Tplus_chan}
\end{figure}

Application of the patching procedure outlined in \S\ref{sec:matching} yields the temperature profiles depicted 
in figure~\ref{fig:Tplus_chan}. 
In the case of symmetric heating (left), deviations from a logarithmic behavior are small,
and the compound temperature law fits well the DNS up to the channel centreline.
Deviations from the logarithmic behavior are on the other hand quite strong in the case
of asymmetric heating.
Again, the agreement with the DNS data is quite good, with observable deviations limited to 
the lowest Prandtl number case.
The correlation provided by \citet{kader_81} also demonstrates reasonable performance in this scenario. However, it exhibits unnatural behavior in the buffer layer and tends to overpredict the strength of the wake.

\subsection{Reynolds number effects} \label{sec:pipe_Re}

\begin{figure}
 \centerline{
 (a)~\includegraphics[width=7.0cm]{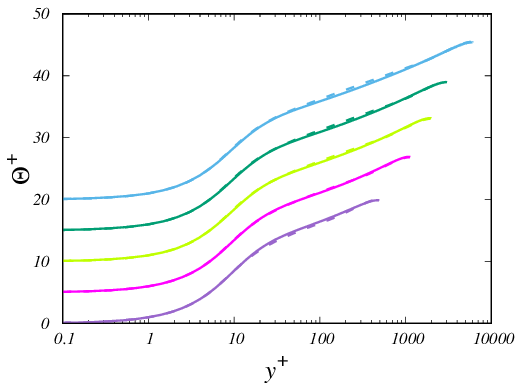}~(b)~\includegraphics[width=7.0cm]{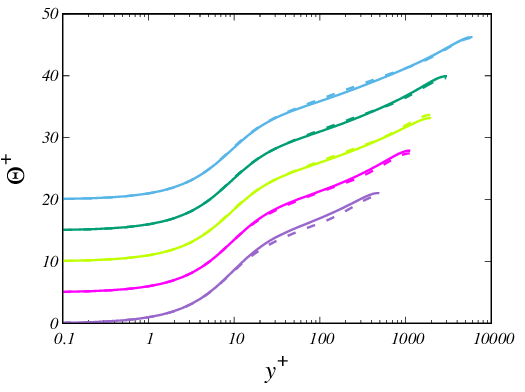}}
\caption{Comparison of mean temperature profiles obtained from DNS (solid lines) 
with the prediction of the compound fitting function given by \eqref{eq:Tprof} and \eqref{eq:parabola}, 
with fitting parameters listed in table~\ref{tab:params} (dashed lines),
	for pipe flow with uniform internal heating (a) and with constant heat flux (b), at $\Pran=1$, for $\Rey_{\tau} = 495, 1132, 1979, 3031, 6019$, from bottom to top, with offset of five wall units between consecutive values.
	}
\label{fig:Tplus_pipe_Re}
\end{figure}

	Assessment of the Reynolds number dependence of our predictions is carried out in this section for the case of pipe flow, for which a wide range of Reynolds numbers has been explored~\citep{pirozzoli_22}, at unit value of the Prandtl number. The DNS data are compared with the analytical predictions herein developed in figure~\ref{fig:Tplus_pipe_Re}. Good agreement of the two distributions is generally found, which further supports universality of our predictions. However, some deviations are observed especially for cases with the lowest Reynolds number, and especially for the case of constant heat flux (panel (b)). This is not unexpected, as the logarithmic law of the wall for the mean temperature, which we leverage on, is known to apply only at sufficiently high Reynolds number that a sizeable overlap emerges between the inner and the outer part of the thermal layer. Further validation of our prediction would involve considering simultaneous Reynolds and Prandtl number variations. To date, no DNS database exists which densely covers the $\Rey-\Pran$ parameter space. However, having established universality of the inner layer with respect to Reynolds number variations, and universality of the outer layer with respect to Prandtl number variations, we see no reason why significant deviations from what herein observed for fixed value of the Reynolds number (Section~\ref{sec:pipe}) and for fixed value of the Prandtl number (present Section) should not apply to arbitrary combination of the two parameters.

\section{Conclusions}

We have presented fully explicit approximations for the mean temperature (or concentration) profiles for forced convection in 
internal flows. The model relies on universality of the inner-scaled mean temperature profile 
in the inner wall layer with respect to Reynolds number variations, and universality of the mean defect 
temperature profile in the core layer with respect to both Reynolds and Prandtl variations.
The explicit representation of the temperature profile \eqref{eq:Tprof} is used for the inner layer,
which effectively incorporates effects from Prandtl number variation. A simple quadratic profile \eqref{eq:parabola} is
then used for the core temperature profile, which includes a single flow-adjustable constant, 
which accounts for geometric and/or thermal forcing effects of the temperature field away from the wall.
A patching condition is derived which guarantees smooth transition between the inner and the core layer.
Comparison with the DNS results shows excellent predictive capability of the model, in a wide range of 
Prandtl numbers, deviations becoming visible only at $\Pran \lesssim 0.1$. As for the effects of 
Reynolds number variation, we find that the model predictions are accurate as long as a sensible
logarithmic layer is present, which occurs when $\Pe_{\tau} = \Pran \, \Rey_{\tau} \gtrsim 11$~\citep{pirozzoli_23b}.
We find that the present model constitutes a significant improvement over previous
models developed for the prediction of the mean temperature distributions~\citep{kader_81}, 
as it yields much more realistic distributions within the wall layer as well as more accurate 
prediction of the shift of the log law with the Prandtl number. 
Given its modularity, the model lends itself to straightforward extension to cases with different flow
geometry (e.g. boundary layers), and/or different heating conditions.\\

DNS data is available at \url{http://newton.dma.uniroma1.it/}. Synthetic temperature profiles and heat transfer coefficients can be generated using ThermoTurb web tool at \url{http://www.thermoturb.com}.

{\bf Acknowledgements} \\
We acknowledge that the results reported in this paper have been achieved using the EuroHPC Research Infrastructure resource LEONARDO based at CINECA, Casalecchio di Reno, Italy, under a LEAP grant.

\bibliography{references}

\end{document}